
\catcode`@=11

\newskip\ttglue

\font\twelverm=cmr12 \font\twelvebf=cmbx12
\font\twelveit=cmti12 \font\twelvesl=cmsl12

\font\ninerm=cmr9
\font\eightrm=cmr8
\font\sixrm=cmr6
\font\eighti=cmmi8   \skewchar\eighti='177
\font\sixi=cmmi6     \skewchar\sixi='177
\font\ninesy=cmsy9   \skewchar\ninesy='60
\font\eightsy=cmsy8  \skewchar\eightsy='60
\font\sixsy=cmsy6    \skewchar\sixsy='60
\font\eightbf=cmbx8
\font\sixbf=cmbx6
\font\eighttt=cmtt8  \hyphenchar\eighttt=-1
\font\eightit=cmti8
\font\eightsl=cmsl8

\def\smalltype{\def\rm{\fam0\eightrm}
 			\textfont0=\eightrm  \scriptfont0=\sixrm  \scriptscriptfont0=\fiverm
 			\textfont1=\eighti   \scriptfont1=\sixi   \scriptscriptfont1=\fivei
 			\textfont2=\eightsy  \scriptfont2=\sixsy  \scriptscriptfont2=\fivesy
 			\textfont3=\tenex    \scriptfont3=\tenex  \scriptscriptfont3=\tenex
    \textfont\itfam=\eightit  \def\it{\fam\itfam\eightit}
	   \textfont\slfam=\eightsl  \def\sl{\fam\slfam\eightsl}
	   \textfont\ttfam=\eighttt  \def\tt{\fam\ttfam\eighttt}
    \textfont\bffam=\eightbf  \scriptfont\bffam=\sixbf
        \scriptscriptfont\bffam=\fivebf  \def\bf{\fam\bffam\eightbf}
    \tt  \ttglue=.5em plus.25em minus.15em
    \normalbaselineskip=9pt
    \setbox\strutbox=\hbox{\vrule height7pt depth2pt width0pt}
    \let\sc=\sixrm  \let\big=\eightbig  \normalbaselines\rm}
\def\eightbig#1{{\hbox{$\textfont0=\ninerm\textfont2=\ninesy
    \left#1\vbox to6.5pt{}\right.\n@space$}}}

\def\medtype{\def\rm{\fam0\tenrm}
 			\textfont0=\tenrm  \scriptfont0=\sevenrm  \scriptscriptfont0=\fiverm
 			\textfont1=\teni   \scriptfont1=\seveni   \scriptscriptfont1=\fivei
 			\textfont2=\tensy  \scriptfont2=\sevensy  \scriptscriptfont2=\fivesy
 			\textfont3=\tenex    \scriptfont3=\tenex  \scriptscriptfont3=\tenex
    \textfont\itfam=\tenit  \def\it{\fam\itfam\tenit}
	   \textfont\slfam=\tensl  \def\sl{\fam\slfam\tensl}
	   \textfont\ttfam=\tentt  \def\tt{\fam\ttfam\tentt}
    \textfont\bffam=\tenbf  \scriptfont\bffam=\sevenbf
        \scriptscriptfont\bffam=\fivebf  \def\bf{\fam\bffam\tenbf}
    \tt  \ttglue=.5em plus.25em minus.15em
    \normalbaselineskip=12pt
    \setbox\strutbox=\hbox{\vrule height8.5pt depth3.5pt width0pt}
    \let\sc=\eightrm  \let\big=\tenbig  \normalbaselines\rm}

\def\bigtype{\let\rm=\twelverm \let\bf=\twelvebf
\let\it=\twelveit \let\sl=\twelvesl \rm}

\def\footnote#1{\edef\@sf{\spacefactor\the\spacefactor}#1\@sf
    \insert\footins\bgroup\smalltype
    \interlinepenalty100 \let\par=\endgraf
    \leftskip=0pt  \rightskip=0pt
    \splittopskip=10pt plus 1pt minus 1pt \floatingpenalty=20000
  \vskip4pt\noindent\hskip20pt\llap{#1\enspace}
\bgroup\strut\aftergroup\@foot\let\next}
\skip\footins=12pt plus 2pt minus 4pt \dimen\footins=30pc

\def\bigfont{\magnification=1200 \baselineskip=20pt}

\def\a{\alpha}  \def\d{\delta}
\def\e{\epsilon} \def\g{\gamma} 
\def\s{\sigma}

\def\cl#1{\centerline{#1}}
\def\clbf#1{\centerline{\bf #1}}

\def\is#1{{\narrower\smallskip\noindent#1\smallskip}}

\long\def\myname{\medskip
\cl{Kiho Yoon}
\cl{Department of Economics, Korea University}
\cl{145 Anam-ro, Seongbuk-gu, Seoul, Korea 02841}
\cl{ \tt kiho@korea.ac.kr}
\cl{\tt http://econ.korea.ac.kr/\~{ }kiho}
\medskip}

\def\ve{\vfill\eject}

\def\frac#1#2{{#1 \over #2}}
\def\Re{I\!\!R}

\newcount\sectnumber
\def\Section#1{\global\advance\sectnumber by 1 \bigskip
           \noindent{\bigtype {\bf \the\sectnumber  \ \ \ #1}} \medskip}

\def\ass#1{\medskip\noindent {\bf Assumption #1.}}

\def\prop#1{\medskip\noindent {\bf Proposition #1.} \it}
\def\lemma#1{\medskip\noindent {\bf Lemma #1.} \it}
\def\thm#1{\medskip\noindent {\bf Theorem #1.} \it}
\def\corr#1{\medskip\noindent {\bf Corollary #1.} \it}

\def\eg#1{\medskip\noindent {\bf Example #1.}}

\def\ok{\smallskip \rm}

\def\pf{\medskip\noindent Proof: \/}
\def\pfo#1{\medskip\noindent {\bf Proof of #1:\/}}
\def\endpf{\hfill {\it Q.E.D.} \smallskip}

\def\app#1{\bigskip {\bigtype \clbf{Appendix #1}} \medskip}

\newcount\notenumber
\def\note#1{\global\advance\notenumber by 1
            \footnote{$^{\the\notenumber}$}{#1}\tenrm}

\def\ref{\bigskip \centerline{\bf REFERENCES} \medskip}

\def\aer{{\it American Economic Review\/ }}

\def\wp{{\it working paper\/ }}

\def\paper#1#2#3#4#5{\noindent\hangindent=20pt#1 (#2), ``#3,'' #4, #5.\par}
\def\wp#1#2#3#4{\noindent\hangindent=20pt#1 (#2), ``#3,'' #4.\par}
\def\book#1#2#3#4{\noindent\hangindent=20pt#1 (#2), {\it #3,} #4.\par}


\bigfont

\def\th{\theta}
\def\Th{\Theta}

{ \ }

\vskip 1cm

{\bigtype
\clbf{Optimal selling time with evolving private information\footnote*{This research was supported by a Korea University Grant (K2612111).}}
}

\vskip 1cm
\bigskip

\myname

\vskip 0.5cm

\clbf{Abstract}
\is{\baselineskip=12pt We study the problem of when to sell one indivisible object to a buyer whose value from immediate allocation follows a privately observed Markov process. Incomplete information transforms what would otherwise be a standard Markovian optimal stopping problem into a dynamic mechanism design problem. We characterize dynamic incentive compatibility by the envelope formula and integral monotonicity. We first solve the relaxed problem that maximizes the seller's revenue over feasible stopping rules while ignoring integral monotonicity, and then give conditions under which the optimal rule of the relaxed problem is dynamically implementable. Under dynamic single crossing and stochastic monotonicity, the optimal rule of the relaxed problem takes a threshold form, and incomplete information weakly delays sale relative to complete information along every realization. We also give a condition for the optimality of a one-step look-ahead rule. Examples illustrate how the informational state depends on the type process.}
\smallskip

\is{\baselineskip=12pt Keywords: optimal stopping, dynamic mechanism design, dynamic pricing, revenue management, threshold rule}
\smallskip

\is{\baselineskip=12pt JEL Classification: C73; D42; D82}

\ve

\Section{Introduction}

We study when a monopolistic seller should sell one indivisible object to a buyer whose value from immediate allocation follows a privately observed Markov process. The object may be an exclusive technology license, a development right, or a specialized productive asset whose value to a particular buyer changes over time. The seller commits to a dynamic mechanism and chooses the sale period to maximize her revenue.

If the buyer's current value were publicly observable, the seller would face a standard Markovian optimal stopping problem. She would compare the current value from selling the object with the option value of retaining it. Under incomplete information, however, the seller must elicit the buyer's value in every period and provide incentives for truthful reporting. Her problem is therefore an optimal stopping problem subject to the dynamic incentive compatibility and participation constraints.

After presenting the complete-information benchmark in Section 2, we formulate the incomplete-information problem as a dynamic mechanism design problem in Section 3. We characterize dynamic incentive compatibility by the envelope formula and integral monotonicity, and express the seller's revenue in terms of dynamic virtual valuations. As is standard in the mechanism design literature, we first solve the relaxed problem while ignoring integral monotonicity and then check whether its solution satisfies the ignored constraint.

The dynamic virtual valuation in period $t$ can be written as
$$\psi_t=\th_t-r_t,$$
where $\th_t$ is the buyer's value for the object and $r_t$ is the inherited information rent, namely, the initial information-rent term propagated through the realized impulse responses of the type process. Thus, the relaxed problem is an optimal stopping problem in which both the current value and the inherited information rent are relevant for the decision. Unlike a standard stopping problem, however, the resulting stopping rule must also satisfy the buyer's incentives to report truthfully in every period.

We first establish a threshold rule for the relaxed problem. Under dynamic single crossing and stochastic monotonicity, the seller sells when the current value is sufficiently high. The threshold is higher when the inherited information rent is larger. The complete-information problem corresponds to the case in which the inherited information rent is zero, and hence incomplete information weakly raises the selling threshold. This comparison is a net result. Incomplete information lowers the revenue from selling in the current period, which favors delay. It also lowers the future revenues available after waiting and hence the value of continuation, which works in the opposite direction. Under the maintained conditions, the first effect dominates. The comparison of thresholds also gives a comparison of sale periods: incomplete information weakly delays sale relative to the efficient complete-information stopping time along every realization.

We next examine dynamic implementability. A sale in the current period eliminates every subsequent allocation. Thus, a higher report may induce an earlier sale and thereby reduce all future allocations to zero. The usual requirement that future allocations be increasing in earlier reports is therefore inappropriate for this stopping problem. We provide an endogenous test for incentive compatibility and more readily verifiable sufficient conditions based on on-path single crossing, a discounted impulse bound, and monotone information rent propagation. The latter conditions are sufficient but not necessary. We also give a closure condition under which a one-step look-ahead rule is optimal.

We present several examples in Section 4. With independent values, the inherited information rent disappears after the first period. With a constant-impulse autoregressive process, it evolves deterministically conditional on the initial report. With a scaling process, the realized impulse responses are state dependent but telescope. Two correlated two-period examples illustrate why implementation must be studied separately from the relaxed stopping problem and why the sufficient implementation conditions are not necessary.

The dynamic mechanism design analysis builds on Pavan {\it et al.} (2014), who derive the envelope and revenue formulas for general dynamic environments. We apply these tools to one irreversible allocation, where a sale in the current period eliminates every future allocation opportunity. This feasibility constraint produces the stopping problem and the implementation issue studied in this paper. We provide a self-contained derivation for general Markovian environments in Appendix A.

This paper is broadly related to dynamic screening and dynamic revenue maximization. Earlier work studies multiperiod adverse selection, sequential screening, and general dynamic mechanisms with evolving private information (Besanko, 1985; Courty and Li, 2000; Battaglini, 2005; Kakade {\it et al.\/}, 2013; Bergemann and Strack, 2015; Bergemann and V\"alim\"aki, 2019). These papers provide important foundations and neighboring applications, but the present paper focuses on one irreversible allocation: sale in the current period removes every future allocation opportunity.

Several papers combine private information with stopping but differ in the timing of screening, the control of stopping, or the available instruments. Board (2007) screens the buyer's private information when an option is initially sold and then delegates the exercise decision through an upfront payment and a strike price.\note{Arve and Zwart (2023) study a related procurement problem in continuous time. In their one-firm case, initial screening is likewise followed by delegated exercise of an investment option as the supplier's privately observed cost evolves. This gives their one-firm problem a real-option structure close to Board (2007).} Kruse and Strack (2015) take a cutoff rule as given and characterize when an informed agent can be induced to follow it using transfers contingent only on the stopping time. In the present paper, the seller receives the buyer's report in every period before sale and chooses the optimal stopping time.\note{A broader dynamic-selling literature studies sequential screening, search and option contracts, learned values for several products, durable-good sales under limited commitment, and changing populations with arrivals, deadlines, or strategic waiting (Gallien, 2006; Pai and Vohra, 2013; Akan {\it et al.\/}, 2015; Mierendorff, 2016; Board and Skrzypacz, 2016; Garrett, 2016; Gershkov {\it et al.\/}, 2018; Ashlagi {\it et al.\/}, 2023; Doval and Skreta, 2024; Hu and Xiao, 2025; Dilm\'e and Garrett, 2026; Krasikov and Lamba, 2026).}

\ve
As far as we know, this is the first paper that studies the revenue-maximizing sale period of one indivisible object to a buyer whose value for the object follows a general privately observed Markov process and characterizes both the optimal stopping structure of the relaxed problem and the dynamic implementability of its solution.

The remainder of the paper is organized as follows. Section 2 presents the complete-information stopping benchmark. Section 3 gives the selling-specific dynamic mechanism formulation, derives the relaxed stopping problem, and studies threshold optimality, the comparison with complete information, dynamic implementation, and the one-step look-ahead rule. Section 4 presents the examples. Section 5 discusses the main findings and directions for further work. Appendix A gives a self-contained derivation of dynamic incentive compatibility and the revenue formula for general Markovian environments and explains how these results relate to Pavan {\it et al.\/} (2014). Appendix B contains the proofs for Section 3, and Appendix C gives the calculations for the final example.

\Section{The complete-information stopping benchmark}

The seller owns one indivisible object and decides when and at what price to sell it to maximize her revenue. There is one buyer. We first study the case where the buyer's current value is publicly observed in every period. We refer to this as the complete-information benchmark; future values remain stochastic and are not known in advance. This is a standard optimal stopping problem for a Markov payoff process (Chow {\it et al.\/}, 1971; Ferguson, 2006; Peskir and Shiryaev, 2006); familiar economic applications include job search (McCall, 1965; Lippman and McCall, 1976). Let $t\in\{1,\ldots,T\}$ denote a period, where $T$ may be infinite, and let $\d\in[0,1]$ be the discount factor.\note{When $T=\infty$, we assume $\d<1$, the relevant discounted payoff series converge absolutely, and the value functions are finite; these assumptions justify the limiting operations and interchanges used below.} Let $X_t\in\Th \subseteq \Re$ be a Markov process with transition distribution $F_{t+1}(\cdot\mid x)$. If the seller transfers the object in period $t$, she can extract the current value $X_t$ and the problem ends. She may also leave the object unallocated. Her problem is
$$\sup_{\tau\leq T}E[\d^{\tau-1}X_\tau],$$
where the supremum also permits no sale, which gives zero revenue. Throughout the paper, indifference between immediate sale and continuation is resolved in favor of immediate sale. Equivalently, we select the earliest optimal stopping time.

Let $V_t^0(x)$ denote the maximal expected payoff from period $t$ onward, evaluated in period $t$. When $T<\infty$, the terminal condition is
$$V_T^0(x)=\max\{x,0\}.$$
For every $t<T$ when $T<\infty$, and for every $t$ when $T=\infty$, the Bellman equation is
$$V_t^0(x)=\max\left\{x,\;\d\int V_{t+1}^0(y)dF_{t+1}(y\mid x)\right\}.$$
In every period with a continuation option, define the complete-information continuation value and stopping gain by
$$C_t^0(x)=\d\int V_{t+1}^0(y)dF_{t+1}(y\mid x), \qquad G_t^0(x)=x-C_t^0(x).$$
When $T < \infty$, set $C_T^0(x)=0$ and $G_T^0(x)=x$. The seller stops when $G_t^0(x)\geq0$.

\prop1 If $G_t^0$ is increasing, the complete-information stopping set is an upper set. Hence there is a cutoff $c_t^0$ such that the seller transfers the object for $x\geq c_t^0$ and waits for $x<c_t^0$. If $c_t^0$ lies in the interior of $\Th$ and $G_t^0$ is continuous at $c_t^0$, then $c_t^0$ satisfies
$$c_t^0=\d\int V_{t+1}^0(y)dF_{t+1}(y\mid c_t^0).$$
\ok

The threshold need not exist for an arbitrary Markov payoff process; it follows from monotonicity or single-crossing conditions. Section 3 imposes analogous conditions on an augmented state. In that formulation, the complete-information benchmark is exactly the zero information rent state:
$$V_t^0(x)=V_t(x,0),\qquad C_t^0(x)=C_t(x,0),\qquad G_t^0(x)=G_t(x,0),\qquad c_t^0=c_t(0).$$
This relation separates the ordinary option value of waiting from the additional effects of incomplete information.

\Section{Optimal stopping with evolving private information}

\noindent {\sl 3.1. Dynamic mechanism and incentive compatibility}
\medskip

The seller owns one indivisible object and decides when and at what price to sell it to maximize her revenue. There is one buyer. Let $t \in \{1, \ldots, T \}$ denote a period, where $T$ may be infinite, and let $\d \in [0,1]$ be the discount factor.\note{When $T=\infty$, the conditions in the previous section are maintained.} The buyer's type in period $t$, which is private information, is $\th_t\in\Th=[\underline\th,\overline\th]$ with $0 \leq \underline \th < \overline \th \leq \infty$. The buyer's type is his value for the object. If the object is not sold, the next type is drawn from $F_{t+1}(\cdot\mid\th_t)$, with density $f_{t+1}(\cdot\mid\th_t)$.

A dynamic (direct) mechanism along the continuation path consists of an allocation rule $q_t: \Th^{t} \rightarrow \{0,1\}$ and a transfer rule $\mu_t: \Th^{t} \rightarrow \Re$ for $t=1, \ldots, T$. Thus, $q_t(\hat \th^t)$ and $\mu_t(\hat \th^t)$ are the action chosen and the transfer, respectively, in period $t$ when the buyer's reports in previous periods are $\hat \th^{t-1}=(\hat \th_1, \ldots, \hat \th_{t-1})$ and the report in this period is $\hat \th_t$.\note{Please refer to Appendix A for a detailed formulation.} Note that, since the object can be allocated at most once, $q_t=1$ ends the mechanism.

Given past reports $\hat\th^{t-1}$, let $U_t(\hat\th_t,\th_t;\hat\th^{t-1})$ be the buyer's continuation payoff when his current type is $\th_t$, his current report is $\hat\th_t$, and he reports truthfully thereafter. With a slight abuse of notation, let $U_t(\th_t;\hat\th^{t-1})=U_t(\th_t,\th_t;\hat\th^{t-1})$. In recursive form,
$$U_t(\hat\th_t,\th_t;\hat\th^{t-1})= \th_t q_t(\hat \th^t) -\mu_t(\hat \th^t) + \d\int_{\underline\th}^{\overline\th} U_{t+1}(y;\hat\th^t)dF_{t+1}(y\mid\th_t).$$
Dynamic incentive compatibility requires
$$U_t(\th_t;\hat\th^{t-1}) \geq U_t(\hat\th_t,\th_t;\hat\th^{t-1}) \quad\hbox{for all }t,\th_t,\hat\th_t,\hat\th^{t-1},$$
and period-1 (interim) participation requires
$$U_1(\th_1)\geq0\qquad\hbox{for every }\th_1. \eqno(1)$$
Define recursively
$$D_t(\hat\th_t,\th_t;\hat\th^{t-1})=\ q_t(\hat \th^t)\ - \d\int_{\underline\th}^{\overline\th}D_{t+1}(y;\hat\th^t)\frac{\partial F_{t+1}(y\mid\th_t)}{\partial\th_t}\,dy, \eqno(2)$$
where we define $D_t(\th_t;\hat\th^{t-1})=D_t(\th_t,\th_t;\hat\th^{t-1})$. The first term is the direct effect in this period and the second term is the indirect effect in the subsequent periods. Dynamic incentive compatibility is characterized by the envelope formula
$$U_t(\th_t;\hat\th^{t-1})=U_t(\underline\th;\hat\th^{t-1})+\int_{\underline\th}^{\th_t}D_t(x;\hat\th^{t-1})dx \eqno(3)$$
and integral monotonicity
$$\int_{\hat\th_t}^{\th_t}D_t(x;\hat\th^{t-1})dx \geq \int_{\hat\th_t}^{\th_t}D_t(\hat\th_t,x;\hat\th^{t-1})dx. \eqno(4)$$
Theorem 3 of Pavan {\it et al.\/} (2014) establishes this characterization for general Markovian environments. In Appendix A, we provide a self-contained derivation. A convenient sufficient condition for (4) is that $D_t(\hat\th_t,\th_t;\hat\th^{t-1})$ be increasing in the current report $\hat\th_t$ for every true type and past history. Pavan {\it et al.\/} (2014) call this average monotonicity.

We impose first-order stochastic dominance throughout this section:
$${\partial F_{t+1}(y\mid x)\over\partial x}\leq0. \eqno(5)$$
This condition implies from (2) that $D_t\geq0$ for every nonnegative stopping allocation. Hence initial utility is increasing, and revenue maximization makes the participation constraint bind at the lowest initial type:
$$U_1(\underline\th)=0. \eqno(6)$$

\medskip
\noindent {\sl 3.2. Dynamic virtual valuations and the relaxed problem}
\medskip

The expected value of the seller's revenue $\sum_{t=1}^T \d^{t-1} \mu_t(\th^t)$ is equal to the expected value of $\sum_{t=1}^T \d^{t-1} \th_t q_t(\th^t)$ minus the expected value of $U_1(\th_1)$. By applying the envelope formula (3) to $U_1(\th_1)$ and recursively expanding $D_t(x;\hat \th^{t-1})$, we get the expected value of the seller's revenue as\note{Please refer to (A6) in Appendix A for a detailed derivation for general Markovian environments, from which we can easily obtain the expression for our selling problem.}
$$E\left[\sum_{t=1}^T\d^{t-1}\psi_t(\th^t)q_t(\th^t)\right]-U_1(\underline\th), \eqno(7)$$
where we define
$$\lambda(x)={1-F_1(x)\over f_1(x)} \eqno(8)$$
and
$$\psi_1(\th_1)=\th_1-\lambda(\th_1), \eqno(9)$$
and, for $t\geq2$ along a no-sale history,
$$\psi_t(\th^t) = \th_t+(-1)^t\lambda(\th_1)\prod_{s=2}^t{{\partial F_s(\th_s\mid\th_{s-1})/\partial\th_{s-1}}
\over f_s(\th_s\mid\th_{s-1})}. \eqno(10)$$
Observe that $\psi_t(\th^t)$ corresponds to the virtual valuation of Myerson (1981) in static mechanism design. Observe also that the seller's revenue does not depend on the transfer rule and thus the revenue equivalence principle applies. The seller's problem is to choose $\{q_t\}_{t=1}^{T}$ for the maximization of this revenue subject to the participation constraint (1), the integral monotonicity constraint (4), and the feasibility constraint that the object can be sold at most once over the entire horizon, i.e., $0\leq \sum_{t=1}^T q_t \leq 1$.

As is standard in the mechanism design literature, we solve the {\it relaxed\/} problem while ignoring the integral monotonicity constraint (4), and then check whether the solution satisfies the ignored constraint. By (6), the relaxed problem chooses a feasible stopping allocation to solve
$$\sup_{\{q_t\}\in{\cal Q}} E\left[\sum_{t=1}^T\d^{t-1}\psi_t(\th^t)q_t(\th^t)\right],$$
where ${\cal Q}$ is the set of allocation rules that satisfy $0\leq \sum_{t=1}^T q_t \leq 1$.

Define the one-step impulse response
$$I_{t+1}(y\mid x) = -{{\partial F_{t+1}(y\mid x)/\partial x}\over f_{t+1}(y\mid x)}\geq0, \eqno(11)$$
where the density is positive, and define the inherited information rent recursively by
$$r_1=\lambda(\th_1),\qquad r_{t+1}=r_tI_{t+1}(\th_{t+1}\mid\th_t). \eqno(12)$$
Then the dynamic virtual valuation has the simple form
$$\psi_t=\th_t-r_t. \eqno(13)$$
Thus, $r_t$ is the initial information-rent term inherited through the realized impulse responses of the type process.

The pair $S_t=(\th_t,r_t)$ is a Markov state. Conditional on $S_t=(x,r)$ and continuation, the next type $Y$ has distribution $F_{t+1}(\cdot\mid x)$ and the next inherited information rent is
$$R=rI_{t+1}(Y\mid x).$$
For every measurable set $B$, let $P_{t+1}(B\mid x,r)$ denote the augmented transition kernel,
$$P_{t+1}(B\mid x,r)=\Pr\{(Y,rI_{t+1}(Y\mid x))\in B\mid S_t=(x,r)\}.$$
The stopping payoff at state $(x,r)$ is $x-r$.

\medskip
\noindent {\sl 3.3. The Bellman equation}
\medskip

Let $V_t(x,r)$ be the maximal expected revenue from period $t$ onward, evaluated in period $t$, conditional on reaching state $(x,r)$ without a sale. No sale gives zero revenue. When $T < \infty$, the terminal condition is
$$V_T(x,r)=\max\{x-r,0\}.$$
For every $t < T$ when $T < \infty$, and for every $t$ when $T=\infty$, define
$$C_t(x,r)=\d\int V_{t+1}(y,rI_{t+1}(y\mid x))dF_{t+1}(y\mid x).$$
In every period with a continuation option, the Bellman equation is
$$V_t(x,r)=\max\{x-r,C_t(x,r)\}.$$
Define the stopping gain by
$$G_t(x,r) = x-r-C_t(x,r).$$
When $T < \infty$, set $C_T(x,r)=0$ and $G_T(x,r)=x-r$. The seller sells if and only if $G_t(x,r)\geq0$. The value premium over immediate sale is
$$H_t(x,r) = V_t(x,r)-(x-r)= \max\{-G_t(x,r), 0\}. \eqno(14)$$
Thus $V_t=x-r+H_t$, and sale is optimal exactly when $H_t=0$.

\medskip
\noindent {\sl 3.4. A general threshold theorem}
\medskip

Introduce the stopping order on the augmented state space:
$$(x,r)\succeq(x',r')\quad\hbox{if and only if}\quad x\geq x'\ \hbox{and}\ r\leq r'.\eqno(15)$$
A state is higher in this order when the buyer's current value is larger and the inherited information rent is smaller. A function is increasing if it is increasing in this order.

\ass1 (Dynamic single crossing) The one-step advantage
$$A_t(x,r)=x-r-\d\int [y-rI_{t+1}(y\mid x)]dF_{t+1}(y\mid x) \eqno(16)$$
is increasing in $(x,r)$ for every $t<T$.

\ass2 (Stochastic monotonicity) The augmented transition kernel $P_{t+1}$ is stochastically increasing in the order (15): for every bounded increasing function $\eta$,
$$\int \eta(s')P_{t+1}(ds'\mid s)$$
is increasing in $s$.

Assumption 1 says that the gain from selling now rather than being required to sell next period has single crossing in the economically relevant state. In static settings, the single-crossing condition says that the marginal change in payoff from a higher allocation is increasing in private information. In comparison, the dynamic single-crossing condition says that the marginal change in payoff from selling earlier is increasing in private information. Kruse and Strack (2015) impose a similar dynamic single-crossing condition in their analysis of optimal stopping with private information. As they note, this condition is reasonable since, in the dynamic setup, allocations differ in the time dimension and the stopping time describes {\it when\/} the object is to be allocated rather than {\it how much\/} to be allocated. Assumption 2 requires a better current state to generate a stochastically better next augmented state. Ordinary first-order stochastic dominance of the current value alone is not sufficient in general because the current type also changes the next inherited information rent through the impulse response.\note{When the proof below applies Assumption 2 to an integrable but possibly unbounded value-premium function, the bounded-function statement is extended by monotone truncation.}

The following coupling condition is a convenient way to verify Assumption 2 from the transition primitives.\note{All proofs are contained in Appendix B.}

\lemma1 Suppose that the next value can be represented as
$$Y=\varphi_{t+1}(x,U),$$
where $U$ has a distribution independent of $x$ and $\varphi_{t+1}(x,u)$ is increasing in $x$ for every $u$. Suppose also that
$$x\longmapsto I_{t+1}(\varphi_{t+1}(x,u)\mid x)$$
is decreasing for every $u$. Then the augmented transition kernel $P_{t+1}$ is stochastically increasing in the stopping order (15).
\ok

The lemma applies, for example, to the autoregressive representation $Y=\g x+(1-\g)\varepsilon$, for which the impulse response is the constant $\g$, and to the scaling representation $Y=xU$, for which the realized impulse response is $U$.

\thm1 Under Assumptions 1 and 2, the stopping gain $G_t(x,r)$ is increasing in the stopping order for every $t$. Hence the stopping region is an upper set. In particular, for every $t$ and $r$ there is a threshold $c_t(r)$ such that the seller sells when $x\geq c_t(r)$ and waits when $x<c_t(r)$. Moreover, $c_t(r)$ is increasing in $r$.
\ok

The proof uses the value premium $H_t$ in (14). Backward induction for a finite horizon, and finite-horizon approximation followed by passage to the limit for an infinite horizon, show that $H_t$ is decreasing in the stopping order, which makes the stopping gain $G_t$ increasing.

The zero information rent state $r=0$ is the complete-information stopping problem for the same underlying value process. Theorem 1 immediately gives a general comparison.

\corr1 Under Assumptions 1 and 2,
$$c_t(r)\geq c_t^0\qquad\hbox{for every }r\geq0.$$
Thus, the inherited information rent weakly raises the stopping threshold state by state.
\ok

The comparison in Corollary 1 is a net result, not a mechanical consequence of replacing $x$ by $x-r$. Incomplete information lowers the revenue from selling immediately, which favors delay. It also lowers the revenues available after waiting and therefore lowers the continuation value, which favors earlier sale. Theorem 1 shows that, under the maintained order conditions, the first effect dominates in the stopping gain: $G_t(x,r)$ is decreasing in $r$. Expressed in current-value units, the incomplete-information cutoff is therefore weakly higher than the complete-information cutoff.

The cutoff comparison also ranks the stopping times themselves. Couple the two rules to the same realization of the underlying value process, including the hypothetical continuation of that process after either rule has stopped.

\corr2 Let $\tau^0$ and $\tau$ denote the complete-information and incomplete-information stopping times. Under Assumptions 1 and 2,
$$\tau\geq\tau^0\qquad\hbox{for every realization of the underlying value process}.$$
Thus, incomplete information weakly delays sale along every realization. Since the stopping rule under complete information maximizes total surplus, incomplete information weakly delays trade relative to the efficient stopping time. \ok

\medskip
\noindent {\sl 3.5. Dynamic implementation of stopping rules}
\medskip

We next give conditions under which the threshold rule of the relaxed problem satisfies dynamic incentive compatibility. Consider a threshold rule after a no-sale report history $\hat\th^{t-1}$. If the current report induces sale, then the first term in (2) is in effect and is equal to one. If it induces continuation, then the first term is equal to zero and the second term, which is
$$J_t(\hat\th_t,\th_t;\hat\th^{t-1})=\d\int_{\underline \th}^{\overline \th} D_{t+1}(y;\hat\th^t) I_{t+1}(y\mid\th_t)dF_{t+1}(y\mid\th_t), \eqno(17)$$
is in effect. Thus,
$$D_t(\hat\th_t,\th_t;\hat\th^{t-1})=\cases{J_t(\hat\th_t,\th_t;\hat\th^{t-1}),& if the report induces continuation;\cr
 1,& if the report induces sale\cr} \eqno(18)$$

\prop2 Suppose that, after every no-sale history,
$$0\leq J_t(\hat\th_t,\th_t;\hat\th^{t-1})\leq1 \eqno(19)$$
for every continuation report and true type, and that $J_t$ is increasing in the current report over the continuation region. Then the threshold rule is dynamically implementable.
\ok

Condition (19) requires that continuation not amplify the discounted marginal effect of the current type beyond the unit marginal effect of immediate sale. It is natural when the effect of the current type on future types is attenuated by persistence and discounting, although it may also hold when the pointwise discounted impulse bound in (21) fails, as Example 5 demonstrates. The second condition in the proposition says that a higher report that still leads to continuation does not reduce $J_t$. Proposition 2 follows directly from average monotonicity.

We now give verifiable sufficient conditions. Theorem 1 gives a threshold in the current value $x$ when the inherited information rent coordinate $r$ is held fixed. Along an actual report path, however, the current report also determines the current inherited information rent. After a no-sale report history $\hat\th^{t-1}$, define
$$\rho_t(x\mid\hat\th^{t-1})=\cases{\lambda(x),&if $t=1$;\cr
 r_{t-1}I_t(x\mid\hat\th_{t-1}),&if $t\geq2$.\cr} \eqno(20)$$
The relevant stopping gain as a function of the current report is therefore
$$x\longmapsto G_t(x,\rho_t(x\mid\hat\th^{t-1})).$$

\ass3 (On-path single crossing) After every no-sale report history $\hat\th^{t-1}$, the stopping set
$$\{x:G_t(x,\rho_t(x\mid\hat\th^{t-1}))\geq0\}$$
is an upper set in the current report. Equivalently, once the on-path gain is nonnegative at some report, it remains nonnegative at every higher report. The stopping set may be empty or the whole type space.

Assumption 3 holds whenever $\rho_t(x\mid\hat\th^{t-1})$ is decreasing in $x$, since Theorem 1 then makes the on-path gain increasing. It may also hold without monotonicity of the on-path gain, as some examples below show.

Next, assume that discounting dominates the one-period impulse response:
$$\d I_{t+1}(y\mid x)\leq1 \quad\hbox{for all }t,x,y. \eqno(21)$$
Finally, impose monotone information rent propagation. For every no-sale report path and every $s>t$, the inherited information rent $r_s$ in (12) is weakly decreasing in an earlier report $\hat\th_t$, holding the intervening and current reports fixed. Since Theorem 1 makes $c_s(r)$ increasing in $r$, this condition implies that a higher earlier continuation report weakly lowers every later stopping threshold.

\ass4 (Monotone information rent propagation) For each $s>t$, the mapping
$$\hat\th_t \longmapsto \lambda(\hat\th_1)\prod_{j=2}^sI_j(\hat\th_j\mid\hat\th_{j-1})$$
is decreasing along no-sale report histories, holding the other reports fixed.

An increasing hazard rate, together with impulse responses whose products have the corresponding monotonicity, is one convenient way to verify Assumption 4.

\thm2 Under Assumptions 1--4 and condition (21), the optimal stopping rule of the relaxed problem satisfies the conditions of Proposition 2. Thus, it is dynamically implementable. \ok

Assumption 3 first makes the optimal stopping rule of the relaxed problem a threshold rule in the current report, rather than merely a threshold in $x$ for a fixed inherited information rent. The proof then uses the recursive representation of $D_t$ given in (2). Condition (21) implies $0\leq D_t\leq1$. Assumption 4 and the monotonicity of $c_s(r)$ in each later period $s$ imply that a higher earlier continuation report cannot raise a later stopping threshold. Hence, $D_t$ is increasing in the current report over the continuation region.

Theorem 2 is sufficient, not necessary. Even when condition (21), Assumption 3, or Assumption 4 fails, the endogenous conditions of Proposition 2 may still be verified directly.

\medskip
\noindent {\sl 3.6. One-step look-ahead rules}
\medskip

A one-step look-ahead (1-SLA) rule compares the current payoff with the expected payoff from stopping in the next period. In this subsection and the corresponding proof, write $w(s)$ for the stopping payoff at state $s$; thus $w(x,r)=x-r$ in the present model. When $T < \infty$, define
$$B_T=\{s:w(s)\geq0\}$$
and, for $t<T$,
$$B_t=\{s:w(s)\geq0\ \hbox{and}\ w(s)\geq\d E[w(S_{t+1})\mid S_t=s]\}. \eqno(22)$$
When $T = \infty$, define $B_t$ by (22) for every $t$. The 1-SLA rule stops upon entering $B_t$.

\prop3 Suppose the one-step stopping regions are closed under continuation: whenever $S_t\in B_t$, every next state that can occur after waiting belongs to $B_{t+1}$ with probability one. Then the one-step look-ahead (1-SLA) rule is optimal for the relaxed problem.
\ok

Under closure, every next state reached from the current stopping region is itself a state at which the 1-SLA rule stops. The full continuation value therefore coincides with the expected payoff from stopping next period.

\Section{Applications and examples}

The examples below have different purposes. The independent case separates the information rent effect from the ordinary option value of waiting. The autoregressive case makes the inherited information rent state transparent. The scaling case shows that the general theory is not confined to constant impulse responses and also gives an explicit stopping rule. The final two-period examples illustrate the implementation condition: one shows why monotonicity of future allocations in past reports is inappropriate, and the other shows that the discounted impulse bound in Theorem 2 is sufficient but not necessary.

\eg1 (Independent values) Suppose
$$F_{t+1}(y\mid x)=F_{t+1}(y).$$
Then $I_{t+1}=0$ and $r_t=0$ for every $t\geq2$. The continuation problem from period 2 onward is therefore the complete-information stopping problem. When $T < \infty$, let
$$
 K_T=0,\qquad
 K_t=\d\int\max\{y,K_{t+1}\}dF_{t+1}(y),\qquad t=T-1,\ldots,1.
$$
When $T=\infty$, let $K_t=C_t^0$; equivalently, $K_t$ satisfies the recursive equation above for every $t$. The seller sells in period $t\geq2$ if and only if $\th_t\geq K_t$, and sells in period 1 if and only if
$$
 \psi_1(\th_1)\geq K_1.
$$
If $\psi_1$ is increasing, the rule is implementable because $D_t=q_t$. The complete-information first-period rule is $\th_1\geq K_1$. Since $\psi_1(\th_1)\leq\th_1$, incomplete information weakly raises the first-period threshold, while all continuation thresholds are unchanged.

\eg2 (Autoregressive values) The geometric propagation of the inherited information rent in this example is related to the decay of informational distortions in multiperiod adverse-selection models (Besanko, 1985; Battaglini, 2005). Suppose
$$\th_{t+1}=\g\th_t+(1-\g)\e_{t+1},\qquad0\leq\g<1,$$
where $\{\e_{t}\}$ are independently and identically distributed random variables over $\Th = [\underline \th, \overline \th]$. The impulse response is constant with $I_{t+1}=\g$ and $r_{t+1}=\g r_t$. That is,
$$r_t=\g^{t-1}\lambda(\th_1).$$

More can be said in this case. By (13) and the last equality, we have
$$\psi_{t+1} = \th_{t+1}-r_{t+1} = \g(\th_t-r_t)+(1-\g)\e_{t+1} = \g\psi_t+(1-\g)\e_{t+1}.$$
Thus, the dynamic virtual valuation follows the same transition law as the buyer's value.

\prop4 In the autoregressive model, the value function of the relaxed problem depends on $(x,r)$ only through $x-r$. Specifically, there is a one-dimensional value function $W_t$ such that
$$V_t(x,r)=W_t(x-r),$$
where, when $T < \infty$,
$$W_T(z)=\max\{z,0\} \eqno(23)$$
and, for every $t < T$ when $T < \infty$, and for every $t$ when $T = \infty$, 
$$W_t(z)=\max\left\{z,\d E\left[W_{t+1}\bigl(\g z+(1-\g)\e_{t+1}\bigr)\right]\right\}. \eqno(24)$$
Consequently, if $c_t^0$ is the complete-information threshold, then
$$c_t(r)=c_t^0+r. \eqno(25)$$
\ok

Proposition 4 gives an exact decomposition of the stopping threshold in this case. The complete-information threshold $c_t^0$ captures the ordinary option value of waiting, while the inherited information rent shifts the current-value threshold upward one for one. Along a no-sale history,
$$
 c_t(r_t)-c_t^0
 =r_t
 =\g^{t-1}\lambda(\th_1).
$$
Thus, the threshold distortion generated by the buyer's initial private information decays geometrically. This is stronger than the general monotonicity conclusion in Theorem 1: in the autoregressive model, the inherited information rent does not otherwise change the shape of the stopping problem but translates its current-value threshold exactly.

Equation (25) is a statewise threshold holding $r$ fixed. Along the report path, the period-1 rule is $\th_1-\lambda(\th_1)\geq c_1^0,$ while, conditional on the initial report, the rule in period $t\geq2$ is $\th_t\geq c_t^0+\g^{t-1}\lambda(\th_1)$.\note{We add that the same argument applies to time-varying affine transitions
$\th_{t+1}=\g_{t+1}\th_t+\eta_{t+1}$, where $\eta_{t+1}$ is independent of the current value. In that case, $r_t=\lambda(\th_1)\prod_{s=2}^t\g_s$, and the current-value threshold remains $c_t^0+r_t$.}

Using the same $\e_{t}$'s to couple two current states shows that the augmented transition is stochastically increasing in the stopping order. Moreover,
$$
 A_t(x,r)=(1-\d\g)(x-r)-\d(1-\g)E[\e_{t+1}],
$$
which is increasing in that order. Theorem 1 therefore gives a threshold $c_t(r)$ that is increasing in the inherited information rent. If $\lambda$ is decreasing, Assumptions 3 and 4 hold, while condition (21) holds automatically since $\d\g<1$. Theorem 2 then gives implementation of the threshold rule of the relaxed problem.

The geometric decay of $r_t$ is the familiar vanishing-distortions property. Its role here is more specific: as $r_t$ falls, the incomplete-information threshold approaches the zero information rent stopping threshold. When the closure condition of Proposition 3 holds, the one-step boundary is
$$x\geq r+{\d(1-\g)E[\e_{t+1}]\over1-\d\g},$$
together with $x-r\geq0$.

\eg3 (Uniform scaling and an explicit stopping rule) Suppose
$$F_{t+1}(y\mid x)={y\over x},\qquad 0\leq y\leq x.$$
Writing $Y=xU$ with $U$ uniform on $[0,1]$, we have
$$
 I_{t+1}(Y\mid x)=U,\qquad
 (Y,R)=(xU,rU).
$$
The common-$U$ coupling makes the augmented kernel stochastically increasing. In addition,
$$
 A_t(x,r)=\left(1-{\d\over2}\right)(x-r),
$$
so Theorem 1 applies. Since $\d U\leq1$, condition (21) holds. Along every report path the product of impulse responses telescopes:
$$
 r_s={\lambda(\hat\th_1)\over\hat\th_1}\hat\th_s.
$$
Thus Assumption 4 holds whenever $\lambda(x)/x$ is decreasing. For $t\geq2$, homogeneity makes the on-path stopping decision independent of the scale of the current report, so the stopping set is either empty or the whole current-report space and Assumption 3 holds. In the uniform initial case below, the explicit first-period rule verifies Assumption 3 when $t=1$. Theorem 2 therefore gives implementation in that case.

For the uniform initial distribution, $\lambda(x)=1-x$, and
$$
 \psi_t={\th_t\over\th_1}(2\th_1-1).
$$
Along a continuation path, $\psi_{t+1}=U\psi_t$. If $\psi_t\geq0$, immediate sale weakly dominates every discounted future revenue path by path and is selected; if $\psi_t<0$, all future revenues are negative and the seller never sells. Thus, it is optimal to sell immediately when $\th_1\geq1/2$ and never sell when $\th_1<1/2$. This example shows that state-dependent impulse responses can nevertheless produce a very simple stopping rule.

\eg4 (Earlier sale truncates future allocation) Let $T=2$, let $F_1(x)=x$ on $[0,1]$, and let
$$
 F_2(y\mid x)=y-2\left(x-{1\over2}\right)y(1-y),\qquad0\leq y\leq1.
$$
This is the transition used in Example 6 of Pavan {\it et al.\/} (2014). We have $\psi_1(x)=2x-1$ and
$$\psi_2(x,y)={2xy^2\over f_2(y\mid x)}\geq0$$
since $f_2(y\mid x)-2(1-x)(1-y)=2xy$, with equality only at the boundary. Consequently,
$$E[\psi_2(x,Y)\mid x]=\int_0^1 2xy^2dy={2x\over3}.$$
Thus, it is optimal to sell in period 1 if and only if
$$x\geq{3\over6-2\d},$$
and otherwise wait and sell in period 2 for every $Y\in[0,1]$.

The second-period allocation is not increasing in the initial report: a sufficiently high report induces sale in period 1 and reduces every future allocation to zero. Thus, a condition requiring future allocations to increase in all past reports fails for the very reason that this is a stopping problem. Nevertheless, we have
$$
 J_1(\hat x,x)
 =-\d\int_0^1{\partial F_2(y\mid x)\over\partial x}dy
 ={\d\over3}
$$
for a continuation report. It is independent of the report and is no larger than one. Proposition 2 therefore gives implementability. This example illustrates why $D_t$, rather than $q_{t+1}$ itself, is appropriate in establishing incentive compatibility.

\eg5 (Endogenous implementation beyond the discounted impulse bound) Let $T=2$, $\d=1$, $F_1(x)=x$, and
$$F_2(y\mid x)=y^x.$$
The period-2 threshold is
$$c_2(x)=\exp\{-x/(1-x)\},$$
and the first-period cutoff is approximately $.6267$. The pointwise impulse response is
$$I_2(y\mid x)=-{y\ln y\over x},$$
which can exceed one when $x$ is small, so condition (21) fails. Nevertheless, the stopping set is an upper set in the current report, and $J_1$ is bounded by one and is increasing in the continuation report. Proposition 2 therefore gives implementation. The calculations are in Appendix C. This example marks the distinction between the sufficient conditions of Theorem 2 and the endogenous implementation test of Proposition 2.

Taken together, the examples show a hierarchy. Independence removes the inherited information rent state after the initial period. Autoregression gives a constant impulse response. Uniform scaling gives a state-dependent but telescoping impulse response. The two-period examples show why implementation must be studied separately from threshold optimality and why the sufficient implementation theorem is not exhaustive.

\ve
\Section{Discussion}

We have studied the problem of when to sell one indivisible object to a buyer whose value from immediate allocation evolves according to a privately observed Markov process. Dynamic incentive compatibility replaces the buyer's current value with the dynamic virtual valuation $\th_t-r_t$, where $r_t$ records the part of the virtual-valuation adjustment inherited from earlier private information through the transition of the value process.

We have solved the relaxed problem that maximizes the seller's revenue over feasible stopping rules while ignoring integral monotonicity, and then established conditions under which the optimal rule of the relaxed problem is dynamically implementable. Under dynamic single crossing and stochastic monotonicity, the optimal stopping rule of the relaxed problem takes a threshold form with the threshold increasing in the inherited information rent, and incomplete information weakly delays sale relative to the efficient complete-information stopping time along every realization. We have also provided an endogenous implementation test and primitive sufficient conditions as well as a condition under which a one-step look-ahead rule is optimal.

Several questions remain. The sufficient implementation conditions may be stronger than necessary, and it would be useful to characterize transition kernels for which the endogenous test holds automatically. A formal analysis of seller instruments could identify when posted prices or an initial menu of option-like contracts suffice and when continuing communication is necessary. An extension to several buyers would also require the joint treatment of dynamic incentive compatibility, selection, and cross-buyer feasibility.

\app{A: The dynamic mechanism design formulation}

This appendix provides a self-contained derivation for general Markovian environments. Let $t \in \{1, \ldots, T \}$ denote a period, where $T$ may be infinite. The player's type in period $t$, which is private information, is $\th_t \in \Th = [\underline \th, \overline \th]$ with $0 \leq \underline \th < \overline \th \leq \infty$. After $\th_t$ is realized in period $t$, a public action $a_t \in A$ is determined. In addition, let $m_t \in \Re$ be a monetary transfer from the player in period $t$. Given sequences $(\th_1, \ldots, \th_T)$ of types and $(a_1, \ldots, a_T)$ of actions, together with a sequence $(m_1, \ldots, m_T)$ of monetary transfers, the player's total payoff is
$$\sum_{t=1}^T \d^{t-1} \bigl[ v(\th_t, a_t) - m_t \bigr],$$
where $\d \in [0,1]$ is the discount factor and $v(\cdot)$ is a (one-period) valuation function.\note{When $T=\infty$, we assume $\d<1$ and absolute convergence sufficient to justify the discounted sums, differentiation under the integral sign, and the changes in the order of summation and integration below.} Let $F_1(\th_1)$ denote the distribution of $\th_1$, with $f_1(\th_1)$ being the corresponding density function. Define $\th^t=(\th_1, \ldots, \th_t)$ and $a^t=(a_1, \ldots, a_t)$, and let $F_t(\th_t |\th^{t-1}, a^{t-1})$ denote the conditional distribution of $\th_t$, with $f_t(\th_t |\th^{t-1}, a^{t-1})$ being the corresponding density function. We impose the following {\it Markov property\/} throughout this appendix:
$$F_t(\th_t|\th^{t-1}, a^{t-1}) = F_t(\th_t|\th_{t-1}, a_{t-1}),$$
that is, $F_t$ does not depend on $\th_s$ or $a_s$ for $s=1, \ldots, t-2$.\note{We may alternatively impose the Markov assumption as $F_t(\th_t|\th^{t-1}, a^{t-1}) = F_t(\th_t|\th_{t-1}, a^{t-1})$, i.e., $F_t$ does not depend on $\th_s$ but depends on $a_s$ for $s=1, \ldots, t-2$. This alternative assumption does not affect the following results.}

A dynamic (direct) mechanism is given by the decision rule $\a_t: \Th^{t-1} \times A^{t-1} \times \Th \rightarrow A$ and the transfer scheme $\mu_t: \Th^{t-1} \times A^{t-1} \times \Th \rightarrow \Re$ for $t=1, \ldots, T$. Thus, $\a_t(\hat \th^{t-1}, a^{t-1}, \hat \th_t)$ and $\mu_t(\hat \th^{t-1}, a^{t-1}, \hat \th_t)$ are the action chosen and the transfer, respectively, in period $t$ when the player's reports and the actions chosen in previous periods are $\hat \th^{t-1}$ and $a^{t-1}$, respectively, and the report in this period is $\hat \th_t$.\note{To be precise, we have to define $\a_1:\Th \rightarrow A$ and $\mu_1: \Th \rightarrow \Re$ differently from those for $t \geq 2$, and to have $\a_1(\hat \th_1)$ and $\mu_1(\hat \th_1)$. We chose to dispense with this extra distinction. Similar conventions apply below.} The player's strategy is $\s_t: \Th^{t-1} \times \Th^{t-1} \times A^{t-1} \times \Th \rightarrow \Th$ for $t=1, \ldots, T$. Thus, $\hat \th_t =\s_t(\th^{t-1}, \hat \th^{t-1}, a^{t-1}, \th_t)$ is the report in period $t$ when the type, report, and action in period $s$ for $s=1, \ldots, t-1$ are $\th_s, \hat \th_s$, and $a_s$, respectively, and the type in this period is $\th_t$.

Define a period-$t$ history
$$\hat h^t=(\hat \th_1, \a_1(\hat \th_1), \hat \th_2, \a_2(\hat \th_1,\a_1(\hat \th_1),\hat \th_2), \cdots, \hat \th_{t-1}, \a_{t-1}(\hat \th_1, \a_1(\hat \th_1), \cdots, \hat \th_{t-1})).$$ In recursive form, $\hat h^t=(\hat h^{t-1}, \hat \th_{t-1}, \a_{t-1}(\hat h^{t-1}, \hat \th_{t-1}))$ with $\hat h^1 = \emptyset$. Thus, $\hat h^t$ is the sequence of reports $\hat \th^{t-1}$ and corresponding actions given by the decision rule. Define recursively
$$\eqalign{&U_t(\hat \th_t, \th_t; \hat \th^{t-1}) = \ v(\th_t, \a_t(\hat h^t, \hat \th_t))-\mu_t(\hat h^t, \hat \th_t) \cr
+ & \d \int_{\th_{t+1}=\underline \th}^{\overline \th} U_{t+1}(\tilde \th_{t+1};\hat \th^{t-1},\hat \th_t)dF_{t+1}(\tilde \th_{t+1}|\th_t, \a_t(\hat h^t, \hat \th_t)),}$$
and define with a slight abuse of notation that
$$U_t(\th_t; \hat \th^{t-1})= U_t(\th_t, \th_t;\hat \th^{t-1}).$$
$U_t(\hat \th_t, \th_t; \hat \th^{t-1})$ is the player's continuation payoff in period $t$ when the report and the action chosen in period $s = 1, \ldots, t-1$ are $\hat \th_s$ and $\a_s(\hat h^s, \hat \th_s)$, respectively, and the true type and the report in this period are $\th_t$ and $\hat \th_t$, respectively. Note that $U_t(\hat \th_t, \th_t; \hat \th^{t-1})$ does not depend on $\th^{t-1}$, the true types in periods $1, \ldots, t-1$.\note{Observe that, under the general transition kernel, $U_t$ depends on $\th_1, \ldots, \th_{t-1}$ as well as $\hat \th_1, \ldots, \hat \th_{t-1}$ and we have $U_t(\hat \th_t, \th_t; \th^{t-1}, \hat \th^{t-1}) = v(\th_t, \a_t(\hat h^t, \hat \th_t))-\mu_t(\hat h^t, \hat \th_t) + \d \int_{\th_{t+1}=\underline \th}^{\overline \th} U_{t+1}(\tilde \th_{t+1};\th^t, \hat \th^t)dF_{t+1}(\tilde \th_{t+1}|$ $\th^t, a^t)$. In this case, $F_{t+1}(\tilde \th_{t+1}|\th^t, a^t)$ depends on the true types $\th^t$ and the chosen actions $a^t$ (that depend on the reports $\hat \th^t$). On the other hand, $F_{t+1}$ does not depend on the true types $\th^{t-1}$ under the Markov assumption.
}

Incentive compatibility is
$$U_t(\th_t; \hat \th^{t-1}) \geq U_t(\hat \th_t, \th_t; \hat \th^{t-1}), \eqno(IC_t)$$
for $t = 1, \ldots, T$.

\noindent Observe that $(IC_t)$ is equivalent to
$$\eqalign{&U_t(\th_t; \hat \th^{t-1}) - U_t(\hat \th_t; \hat \th^{t-1}) \geq  v(\th_t, \a_t(\hat h^t,\hat \th_t))-v(\hat \th_t, \a_t(\hat h^t,\hat \th_t)) \cr
+ & \ \d \int_{\th_{t+1}=\underline \th}^{\overline \th} U_{t+1}(\tilde \th_{t+1};\hat \th^{t-1},\hat \th_t)d\bigl(F_{t+1}(\tilde \th_{t+1}|\th_t, \a_t(\hat h^t, \hat \th_t)) - F_{t+1}(\tilde \th_{t+1}|\hat \th_t, \a_t(\hat h^t, \hat \th_t)) \bigr).} \eqno(IC'_t)$$
Interchanging the roles of $\th_t$ and $\hat \th_t$, we have
$$\eqalign{&U_t(\hat \th_t; \hat \th^{t-1}) - U_t(\th_t; \hat \th^{t-1}) \geq  v(\hat \th_t, \a_t(\hat h^t, \th_t))-v(\th_t, \a_t(\hat h^t,\th_t)) \cr
+ & \ \d \int_{\th_{t+1}=\underline \th}^{\overline \th} U_{t+1}(\tilde \th_{t+1};\hat \th^{t-1},\th_t)d\bigl(F_{t+1}(\tilde \th_{t+1}|\hat \th_t, \a_t(\hat h^t, \th_t)) - F_{t+1}(\tilde \th_{t+1}|\th_t, \a_t(\hat h^t, \th_t))\bigr).}$$
Combining these inequalities, we get
$$\eqalign{&v(\hat \th_t, \a_t(\hat h^t,\hat \th_t)) - v(\th_t, \a_t(\hat h^t,\hat \th_t)) \cr
+ & \ \d \int_{\th_{t+1}=\underline \th}^{\overline \th} U_{t+1}(\tilde \th_{t+1};\hat \th^{t-1},\hat \th_t)d\bigl(F_{t+1}(\tilde \th_{t+1}|\hat \th_t, \a_t(\hat h^t, \hat \th_t)) - F_{t+1}(\tilde \th_{t+1}|\th_t, \a_t(\hat h^t, \hat \th_t)) \bigr) \cr
\geq & U_t(\hat \th_t; \hat \th^{t-1}) - U_t(\th_t; \hat \th^{t-1}) \cr
\geq & v(\hat \th_t, \a_t(\hat h^t, \th_t))-v(\th_t, \a_t(\hat h^t,\th_t)) \cr
+ & \ \d \int_{\th_{t+1}=\underline \th}^{\overline \th} U_{t+1}(\tilde \th_{t+1};\hat \th^{t-1},\th_t)d\bigl(F_{t+1}(\tilde \th_{t+1}|\hat \th_t, \a_t(\hat h^t, \th_t)) - F_{t+1}(\tilde \th_{t+1}|\th_t, \a_t(\hat h^t, \th_t))\bigr).}$$

We will assume that both $v(\cdot)$ and $f_t(\cdot)$ for $t=1, \ldots, T$ are continuously differentiable and that $\a_t(\cdot)$ for $t=1, \ldots, T$ is differentiable almost everywhere. Then, dividing the previous inequalities by $\hat \th_t - \th_t$ and taking limits, we get
$$\eqalign{&\frac{dU_t(\th_t; \hat \th^{t-1})}{d\th_t}=v_\th(\th_t, \a_t(\hat h^t, \th_t)) \cr
+& \d \int_{\underline \th}^{\overline \th} U_{t+1}(\tilde \th_{t+1};\hat \th^{t-1}, \th_t) \ \frac{\partial f_{t+1}(\tilde \th_{t+1}|\th_t, \a_t(\hat h^t, \th_t))}{\partial \th_t} d\tilde \th_{t+1}}$$
almost everywhere. Note that the notation $v_\th(\th, a)$ is the partial derivative of $v(\th,a)$ with respect to $\th$ and that $\partial f_{t+1}/\partial \th_t$ is only with respect to $\th_t$ in $f_{t+1}(\th_{t+1}|\th_t, a_t)$. We have
$$\eqalign{\frac{dU_t(\th_t;\hat \th^{t-1})}{d\th_t} = & \ v_\th(\th_t, \a_t(\hat h^t, \th_t)) + \d \Bigl[ U_{t+1}(\tilde \th_{t+1};\hat \th^{t-1}, \th_t) \frac{\partial F_{t+1}(\tilde \th_{t+1}|\th_t, \a_t(\hat h^t, \th_t))}{\partial \th_t} \Bigr]_{\underline \th}^{\overline \th} \cr
-& \d \int_{\underline \th}^{\overline \th} \frac{dU_{t+1}(\tilde \th_{t+1};\hat \th^{t-1}, \th_t)}{d\th_{t+1}} \ \frac{\partial F_{t+1}(\tilde \th_{t+1}|\th_t, \a_t(\hat h^t, \th_t))}{\partial \th_t} d\tilde \th_{t+1} \cr
= \ v_\th(\th_t, \a_t(\hat  h^t, & \th_t))  - \d \int_{\underline \th}^{\overline \th} \frac{dU_{t+1}(\tilde \th_{t+1};\hat \th^{t-1}, \th_t)}{d\th_{t+1}} \ \frac{\partial F_{t+1}(\tilde \th_{t+1}|\th_t, \a_t(\hat h^t, \th_t))}{\partial \th_t} d\tilde \th_{t+1},}$$
where the first equality follows from integration by parts and the second equality follows from the fact that $F_{t+1}(\underline \th|\th_t, \a_t(\hat h^t, \th_t)) = 0$ and $F_{t+1}(\overline \th|\th_t, \a_t(\hat h^t, \th_t)) = 1$ for all $\th_t$ and so $\partial F_{t+1}/\partial \th_t = 0$ when $\tilde \th_{t+1} = \underline \th$ or $\overline \th$.

By recursion, we have

{\smalltype
$$\eqalign{&\frac{dU_t(\th_t;\hat \th^{t-1})}{d\th_t}=v_\th(\th_t, \a_t(\hat h^t, \th_t)) \cr
+ & \sum_{s=t+1}^T (-\d)^{s-t} \underbrace{\int_{\underline \th}^{\overline \th} \cdots \int_{\underline \th}^{\overline \th}}_{(s-t)-{\rm times}} v_\th(\tilde \th_s, \a_s(\hat h^t,\th_t, \a_t(\hat h^t, \th_t),\tilde \th_{t+1}, \a_{t+1}(\hat h^t, \th_t, \a_t(\hat h^t,\th_t),\tilde \th_{t+1}), \cdots, \tilde \th_s)) \cr
\times & \frac{\partial F_{t+1}(\tilde \th_{t+1}|\th_t, \a_t(\hat h^t, \th_t))}{\partial \th_t} \cdots \frac{\partial F_s(\tilde \th_s|\tilde \th_{s-1}, \a_{s-1}(\hat h^t, \th_t, \a_t(\hat h^t,\th_t),\tilde \th_{t+1}, \cdots, \tilde \th_{s-1})) }{\partial \th_{s-1}} d\tilde \th_s \cdots d\tilde \th_{t+1}} \eqno(A1)$$
}
almost everywhere.

Define
$$h_+^s(\hat h^t, \th_t) = (\hat h^t, \th_t, \a_t(\hat h^t, \th_t), \tilde \th_{t+1}, \a_{t+1}(\hat h^t, \th_t, \a_t(\hat h^t, \th_t), \tilde \th_{t+1}), \cdots, \tilde \th_s),$$
that is, $h_+^s(\hat h^t, \th_t)$ is the period-$s$ history plus period-$s$ report $\tilde \th_s$, conditional on period-$t$ history $\hat h^t$ and period-$t$ report $\th_t$. Then, \par
\cl{$h_+^s(\hat h^t, \hat \th_t) = (\hat h^t, \hat \th_t, \a_t(\hat h^t, \hat \th_t), \tilde \th_{t+1}, \a_{t+1}(\hat h^t, \hat \th_t, \a_t(\hat h^t, \hat \th_t), \tilde \th_{t+1}), \cdots, \tilde \th_s)$.}

\noindent Define
$$\eqalign{&D_t(\hat \th_t, \th_t;\hat \th^{t-1})=v_\th(\th_t, \a_t(\hat h^t, \hat \th_t)) \cr
+ & \sum_{s=t+1}^T (-\d)^{s-t} \underbrace{\int_{\underline \th}^{\overline \th} \cdots \int_{\underline \th}^{\overline \th}}_{(s-t)-{\rm times}} v_\th(\tilde \th_s, \a_s(h_+^s(\hat h^t, \hat \th_t))) \ \frac{\partial F_{t+1}(\tilde \th_{t+1}|\th_t, \a_t(\hat h^t, \hat \th_t))}{\partial \th_t} \cr
\times &
\frac{\partial F_{t+2}(\tilde \th_{t+2}|\tilde \th_{t+1}, \a_{t+1}(h_+^{t+1}(\hat h^t, \hat \th_t)))}{\partial \th_{t+1}}
\cdots \frac{\partial F_s(\tilde \th_s|\tilde \th_{s-1}, \a_{s-1}(h_+^{s-1}(\hat h^t, \hat \th_t)))}{\partial \th_{s-1}}d\tilde \th_s \cdots d\tilde \th_{t+1}.} \eqno(A2)$$
Note that this captures the current and future impact of the change in $\th_t$ when the current true type is $\th_t$, the current report is $\hat \th_t$, and the past reports are $\hat \th^{t-1}$. Note also that (A2) can be written in recursive form as
$$\eqalign{D_t(\hat \th_t, \th_t; \hat \th^{t-1})=& \ v_\th(\th_t, \a_t(\hat h^t, \hat \th_t)) \cr
- & \d \int_{\underline \th}^{\overline \th} D_{t+1}(\tilde \th_{t+1};\hat \th^t) \ \frac{\partial F_{t+1}(\tilde \th_{t+1}|\th_t, \a_t(\hat h^t, \hat \th_t))}{\partial \th_t} d\tilde \th_{t+1}.} \eqno(A3)$$
Define with a slight abuse of notation that $D_t(\th_t; \hat \th^{t-1})=D_t(\th_t, \th_t; \hat \th^{t-1})$.
Then, (A1) can be rewritten as
$$\frac{dU_t(\th_t;\hat \th^{t-1})}{d\th_t}=D_t(\th_t; \hat \th^{t-1})$$
almost everywhere. Under the maintained integrability conditions, we have
$$U_t(\th_t; \hat \th^{t-1}) = U_t(\underline \th ; \hat \th^{t-1}) + \int_{\underline \th}^{\th_t} D_t(\tilde \th_t; \hat \th^{t-1}) d\tilde \th_t.$$
In particular, we have
$$U_1(\th_1)=U_1(\underline \th)+\int_{\underline \th}^{\th_1} D_1(\tilde \th_1) d\tilde \th_1,$$
where
$$\eqalign{&D_1(\tilde \th_1) = v_\th(\tilde \th_1, \a_1(\tilde \th_1)) + \sum_{t=2}^T (-\d)^{t-1} \underbrace{\int_{\underline \th}^{\overline \th} \cdots \int_{\underline \th}^{\overline \th}}_{(t-1)-{\rm times}} v_\th(\tilde \th_t, \a_t(\tilde h^t, \tilde \th_t)) \cr
& \times \frac{\partial F_2(\tilde \th_2|\tilde \th_1, \a_1(\tilde \th_1))}{\partial \th_1} \cdots \frac{\partial F_t(\tilde \th_t|\tilde \th_{t-1}, \a_{t-1}(\tilde h^{t-1}, \tilde \th_{t-1}))}{\partial \th_{t-1}} d\tilde \th_t \cdots d\tilde \th_2.}$$

\noindent We now present a necessary and sufficient condition for incentive compatibility.

\thm{A} Incentive compatibility holds if and only if
$$U_t(\th_t; \hat \th^{t-1}) = U_t(\underline \th ; \hat \th^{t-1}) + \int_{\underline \th}^{\th_t} D_t(\tilde \th_t; \hat \th^{t-1}) d\tilde \th_t \ \ \ \forall t, \forall \th_t {\rm \ and \ } \forall \hat \th^{t-1}; \eqno(A4)$$
and
$$\int_{\hat \th_t}^{\th_t} D_t(\tilde \th_t; \hat \th^{t-1}) d\tilde \th_t \geq \int_{\hat \th_t}^{\th_t} D_t(\hat \th_t, \tilde \th_t; \hat \th^{t-1}) d\tilde \th_t \ \ \ \forall t, \forall \th_t, \forall \hat \th_t {\rm \ and \ } \forall \hat \th^{t-1}. \eqno(A5)$$
\ok

\pf ($\Leftarrow$ part) Observe first that, by (A4), the left-hand side of (A5) is equal to $U_t(\th_t;\hat \th^{t-1})-U_t(\hat \th_t; \hat \th^{t-1})$. Observe next that the right-hand side of (A5) is equal to

$$\eqalign{&v(\th_t, \a_t(\hat h^t, \hat \th_t))-v(\hat \th_t, \a_t(\hat h^t, \hat \th_t)) \cr
- & \d \int_{\hat \th_t}^{\th_t} \int_{\underline \th}^{\overline \th} D_{t+1}(\tilde \th_{t+1}; \hat \th^t) \ \frac{\partial F_{t+1}(\tilde \th_{t+1}|\tilde \th_t, \a_t(\hat h^t, \hat \th_t))}{\partial \th_t} d\tilde \th_{t+1} d\tilde \th_t \cr
= \ &v(\th_t, \a_t(\hat h^t, \hat \th_t))-v(\hat \th_t, \a_t(\hat h^t, \hat \th_t)) \cr
- & \d \int_{\hat \th_t}^{\th_t} \int_{\underline \th}^{\overline \th} \frac{dU_{t+1}(\tilde \th_{t+1}; \hat \th^t)}{d\th_{t+1}} \ \frac{\partial F_{t+1}(\tilde \th_{t+1}|\tilde \th_t, \a_t(\hat h^t, \hat \th_t))}{\partial \th_t}  d\tilde \th_{t+1} d\tilde \th_t \cr
= \ &v(\th_t, \a_t(\hat h^t, \hat \th_t))-v(\hat \th_t, \a_t(\hat h^t, \hat \th_t)) \cr
- & \d \int_{\underline \th}^{\overline \th} \frac{dU_{t+1}(\tilde \th_{t+1}; \hat \th^t)}{d\th_{t+1}} \int_{\hat \th_t}^{\th_t} \frac{\partial F_{t+1}(\tilde \th_{t+1}|\tilde \th_t, \a_t(\hat h^t, \hat \th_t))}{\partial \th_t} d\tilde \th_t d\tilde \th_{t+1} \cr
= \ &v(\th_t, \a_t(\hat h^t, \hat \th_t))-v(\hat \th_t, \a_t(\hat h^t, \hat \th_t))  \cr
- & \d \int_{\underline \th}^{\overline \th} \frac{dU_{t+1}(\tilde \th_{t+1}; \hat \th^t)}{d\th_{t+1}} \Bigl(F_{t+1}(\tilde \th_{t+1}|\th_t, \a_t(\hat h^t, \hat \th_t))-F_{t+1}(\tilde \th_{t+1}|\hat \th_t, \a_t(\hat h^t, \hat \th_t))\Bigr)d\tilde \th_{t+1} \cr
= \ &v(\th_t, \a_t(\hat h^t, \hat \th_t))-v(\hat \th_t, \a_t(\hat h^t, \hat \th_t)) \cr
- & \d \Bigl[U_{t+1}(\tilde \th_{t+1};\hat \th^t) \Bigl(F_{t+1}(\tilde \th_{t+1}|\th_t, \a_t(\hat h^t, \hat \th_t))-F_{t+1}(\tilde \th_{t+1}|\hat \th_t, \a_t(\hat h^t, \hat \th_t))\Bigr)\Bigr]_{\underline \th}^{\overline \th} \cr
+ & \d \int_{\underline \th}^{\overline \th} U_{t+1}(\tilde \th_{t+1}; \hat \th^t) d\Bigl( F_{t+1}(\tilde \th_{t+1}|\th_t, \a_t(\hat h^t, \hat \th_t))-F_{t+1}(\tilde \th_{t+1}|\hat \th_t, \a_t(\hat h^t, \hat \th_t)) \Bigr) \cr
= \ &v(\th_t, \a_t(\hat h^t, \hat \th_t))-v(\hat \th_t, \a_t(\hat h^t, \hat \th_t)) \cr
+ & \d \int_{\underline \th}^{\overline \th} U_{t+1}(\tilde \th_{t+1}; \hat \th^t) d\Bigl( F_{t+1}(\tilde \th_{t+1}|\th_t, \a_t(\hat h^t, \hat \th_t))-F_{t+1}(\tilde \th_{t+1}|\hat \th_t, \a_t(\hat h^t, \hat \th_t)) \Bigr).}$$
The first expression follows from (A3), the first equality follows from the differential form of (A4) for $U_{t+1}(\tilde \th_{t+1} ; \hat \th^t)$, the second equality follows from the change in the order of integration, the third equality follows from integrating out the inner integral, the fourth equality follows from integration by parts, and the last equality follows from the fact that $F_{t+1}(\underline \th|\th_t, \a_t(\hat h^t, \hat \th_t)) = F_{t+1}(\underline \th|\hat \th_t, \a_t(\hat h^t, \hat \th_t))=0$ and $F_{t+1}(\overline \th|\th_t, \a_t(\hat h^t, \hat \th_t))=\ F_{t+1}(\overline \th|\hat \th_t,$ $\a_t(\hat h^t, \hat \th_t))=1$. Putting these together, this is nothing but $(IC'_t)$.

\noindent ($\Rightarrow$ part) Recall from the discussion preceding this theorem that $(IC_t)$ implies (A4). It is also straightforward to see that $(IC_t)$ implies (A5): Follow the reverse steps of the previous argument. \endpf

The analysis in this appendix is closely related to the path-breaking work of Pavan {\it et al.\/} (2014). They establish many seminal results for general dynamic environments before delving into Markovian environments. In contrast, we focus on Markovian environments and present the material in a rather elementary fashion. Our analysis hopefully demonstrates that the techniques of static mechanism design can be straightforwardly extended and adapted to dynamic settings. Moreover, it makes the optimal selling problem of the text easy to deal with.

Theorem A above is essentially the same as Theorem 3 of Pavan {\it et al.\/} (2014), which states a necessary and sufficient condition for incentive compatibility under Markovian environments. They call inequality (A5) {\it integral monotonicity.} They also present several sufficient conditions in their Corollary 1 of Theorem 3. They call the condition that $D_t(\hat \th_t, \th_t; \hat \th^{t-1})$ is increasing in $\hat \th_t$ {\it average monotonicity.\/}

By definition of $U_1(\th_1)$, the total expected payment the player makes, i.e., the expected value of $\sum_{t=1}^T \d^{t-1} \mu_t(\tilde h^t, \tilde \th_t)$, is
$$\eqalign{&\int_{\underline \th}^{\overline \th} v(\tilde \th_1, \a_1(\tilde \th_1))f_1(\tilde \th_1)d\tilde \th_1 + \sum_{t=2}^T \d^{t-1} \underbrace{\int_{\underline \th}^{\overline \th} \cdots \int_{\underline \th}^{\overline \th}}_{t-{\rm times}}  v(\tilde \th_t, \a_t(\tilde h^t, \tilde \th_t)) \cr
& \ \times f_t(\tilde \th_t|\tilde \th_{t-1}, \a_{t-1}(\tilde h^{t-1},\tilde \th_{t-1})) \cdots f_2(\tilde \th_2|\tilde \th_1, \a_1(\tilde \th_1)) f_1(\tilde \th_1) d\tilde \th_t \cdots d\tilde \th_1 }$$

$$\hskip -8cm - \int_{\underline \th}^{\overline \th} U_1(\tilde \th_1) f_1(\tilde \th_1)d\tilde \th_1.$$

\noindent Since
$$\eqalign{&\int_{\underline \th}^{\overline \th} U_1(\tilde \th_1) f_1(\tilde \th_1)d\tilde \th_1 = \ \Bigl[-U_1(\tilde \th_1)(1-F_1(\tilde \th_1))\Bigr]_{\underline \th}^{\overline \th} + \int_{\underline \th}^{\overline \th} \frac{dU_1(\tilde \th_1)}{d\th_1} (1-F_1(\tilde \th_1))d\tilde \th_1 \cr
= & \ U_1(\underline \th) + \int_{\underline \th}^{\overline \th} D_1 (\tilde \th_1) (1-F_1(\tilde \th_1)) d\tilde \th_1}$$
where the second equality holds by (A4), the total expected payment is equal to
$$\eqalign{& \int_{\underline \th}^{\overline \th} \Bigl[v(\tilde \th_1, \a_1(\tilde \th_1)) - v_\th(\tilde \th_1, \a_1(\tilde \th_1)) \ \frac{1-F_1(\tilde \th_1)}{f_1(\tilde \th_1)} \Bigr] f_1(\tilde \th_1) d\tilde \th_1 \cr
+ &  \sum_{t=2}^T \d^{t-1} \int_{\underline \th}^{\overline \th} \cdots \int_{\underline \th}^{\overline \th} \Bigl[ v(\tilde \th_t, \a_t(\tilde h^t, \tilde \th_t)) + (-1)^{t} \ v_\th(\tilde \th_t, \a_t(\tilde h^t, \tilde \th_t)) \frac{1-F_1(\tilde \th_1)}{f_1(\tilde \th_1)}  \cr
& \ \ \times \frac{\partial F_2(\tilde \th_2|\tilde \th_1, \a_1(\tilde \th_1))/\partial \th_1}{f_2(\tilde \th_2|\tilde \th_1, \a_1(\tilde \th_1))} \cdots \frac{\partial F_t(\tilde \th_t|\tilde \th_{t-1}, \a_{t-1}(\tilde h^{t-1},\tilde \th_{t-1}))/\partial \th_{t-1}}{f_t(\tilde \th_t|\tilde \th_{t-1}, \a_{t-1}(\tilde h^{t-1},\tilde \th_{t-1}))} \Bigr] \cr
& \ \ \times f_t(\tilde \th_t|\tilde \th_{t-1}, \a_{t-1}(\tilde h^{t-1},\tilde \th_{t-1})) \cdots f_1(\tilde \th_1) d\tilde \th_t \cdots d\tilde \th_1 \cr
- & \  U_1(\underline \th)} \eqno(A6)$$
\noindent by (A2).

\app{B: Proofs for Section 3}

\pfo{Lemma 1} Couple two current states $(x,r)\succeq(x',r')$ using the same realization of $U$. The assumptions give
$$\varphi_{t+1}(x,U)\geq\varphi_{t+1}(x',U)$$
and
$$I_{t+1}(\varphi_{t+1}(x,U)\mid x)\leq I_{t+1}(\varphi_{t+1}(x',U)\mid x').$$
Since $r\leq r'$ and the impulse responses are nonnegative,
$$rI_{t+1}(\varphi_{t+1}(x,U)\mid x)
\leq r'I_{t+1}(\varphi_{t+1}(x',U)\mid x').$$
Thus the coupled next state from $(x,r)$ is weakly higher in the stopping order for every $U$. Taking expectations of any bounded increasing function proves Assumption 2. \endpf

\pfo{Theorem 1} Suppose first that $T < \infty$. Let $H_t$ be defined by (14). Since $G_T(x,r)=x-r$ is increasing in the stopping order, $H_T$ is decreasing. Suppose $H_{t+1}$ is decreasing. If $H_{t+1}$ is unbounded, set $H_{t+1}^n=\min\{H_{t+1},n\}$. Then $H_{t+1}^n$ is bounded and decreasing, so Assumption 2 implies that
$$E[H_{t+1}^n(S_{t+1})\mid S_t=(x,r)]$$
is decreasing in $(x,r)$. Monotone convergence and the maintained finiteness of the value functions extend this conclusion to $H_{t+1}$. Moreover,
$$\eqalign{G_t(x,r)=&\ x-r-\d E[V_{t+1}(S_{t+1})\mid S_t=(x,r)]\cr
=&\ A_t(x,r)-\d E[H_{t+1}(S_{t+1})\mid S_t=(x,r)].}$$
The first term is increasing by Assumption 1 and the negative of the second expectation is increasing by Assumption 2 and the truncation argument. Hence $G_t$ is increasing. It follows from (14) that $H_t$ is decreasing, completing the induction. When $T=\infty$, apply the same recursive argument to the finite-horizon values $V_t^{(N)}$ obtained by allowing sale only through period $N$. The preceding conclusion holds for every $N$. Moreover, $V_t^{(N)}\uparrow V_t$ by the definition of the infinite-horizon stopping value and the maintained finiteness of $V_t$. Hence $H_t^{(N)}=V_t^{(N)}-(x-r)\uparrow H_t$, and the pointwise limit $H_t$ is decreasing. The displayed identity then implies that $G_t$ is increasing. The stopping set $\{(x,r):G_t(x,r)\geq0\}$ is therefore an upper set. Holding $r$ fixed gives the current-value threshold, and holding $x$ fixed shows that the threshold is increasing in $r$. \endpf

\pfo{Corollary 2} Couple the two rules to the same realization of the underlying value process. If the incomplete-information rule stops in period $t$, then
$$\th_t\geq c_t(r_t)\geq c_t(0)=c_t^0,$$
so the complete-information rule also stops in or before period $t$. Thus $\tau^0\leq\tau$ for every realization. Under complete information the seller can extract the buyer's current value, so the complete-information stopping rule maximizes total surplus. \endpf

\pfo{Proposition 2} By (18), $D_t$ is increasing in the report over the continuation region by hypothesis, jumps weakly upward to one upon entering the stopping region by (19), and remains one over the stopping region. Thus $D_t(\hat\th_t,\th_t;\hat\th^{t-1})$ is increasing in $\hat\th_t$. Average monotonicity implies (4). \endpf

\pfo{Theorem 2} By Assumption 3, after every no-sale history the stopping set is an upper set in the current report. Hence the optimal stopping rule has the threshold form required by Proposition 2.

We first show that $0\leq D_t\leq1$ after every reported history. Suppose first that $T<\infty$. In period $T$, we have $D_T=q_T\in\{0,1\}$. Suppose that $0\leq D_{t+1}\leq1$. As discussed in (18), if the current report induces sale then $D_t=1$ whereas if it induces continuation then $D_t=J_t$.
By the first-order stochastic dominance condition (5), we have $I_{t+1}\geq0$; by the induction hypothesis and (21), the integrand is between zero and one. Hence $0\leq D_t\leq1$. This completes the backward induction. When $T=\infty$, truncate the explicit discounted representation of $D_t$ derived in Appendix A after period $N$. Each finite truncation satisfies the same bound, and the maintained convergence implies $0\leq D_t\leq1$ in the limit.

It remains to establish that $D_t$ is increasing in the current report over the continuation region. Fix a no-sale history, a true current type $\th_t$, and two continuation reports $x<x'$. By Assumption 4, holding all subsequent reports fixed, the inherited information rent in every later period is weakly lower following $x'$ than following $x$. Since $c_s(r)$ is increasing in $r$, every subsequent stopping threshold is therefore weakly lower following $x'$.

For a finite horizon, we show by backward induction that $D_s$ is weakly larger in every later period $s$ following $x'$ than following $x$, for every common subsequent report history and true type. In period $T$, a lower threshold weakly raises $q_T$ and so $D_T$ since $D_T=q_T$. Suppose the comparison holds in period $s+1$. If both histories prescribe sale in period $s$, both values of $D_s$ equal one. If the history following $x$ prescribes continuation and the history following $x'$ prescribes sale, $D_s$ rises from a value in $[0,1]$ to one. If both histories prescribe continuation, (2) and the induction hypothesis imply that $D_s$ is weakly larger following $x'$. The case when the history following $x$ prescribes sale and the history following $x'$ prescribes continuation cannot occur because the stopping threshold is weakly lower following $x'$. Hence, the comparison holds in period $s$. In particular, $D_{t+1}(y;\hat\th^{t-1},x) \leq D_{t+1}(y;\hat\th^{t-1},x')$ for every $y$. Since we have chosen both $x$ and $x'$ that induce continuation, (2) gives $D_t(x,\th_t;\hat\th^{t-1}) \leq D_t(x',\th_t;\hat\th^{t-1})$.
When $T=\infty$, the same comparison holds for every finite truncation of the explicit discounted representation of $D_s$ and therefore, by the maintained convergence, for $D_s$ itself.

Thus, after every no-sale history, $D_t$ is bounded between zero and one for every continuation report and true type and is increasing in the current report over the continuation region. Proposition 2 then gives the desired result. \endpf

\pfo{Proposition 3} Suppose first that $T < \infty$. In period $T$, the result follows from the definition of $B_T$. Suppose the 1-SLA rule is optimal from $t+1$ onward. If $s\notin B_t$, either $w(s)<0$ or $w(s)<\d E[w(S_{t+1})\mid S_t=s]$. Since $V_{t+1}\geq\max\{w,0\}$, continuation is better than stopping. If $s\in B_t$, closure implies that every next state reached after waiting lies in $B_{t+1}$. By the induction hypothesis, the rule stops next period at all such states, so $V_{t+1}=w$ there. The continuation value is thus the one-period-ahead expected payoff and is no larger than $w(s)$. Hence, stopping is optimal.

When $T=\infty$, closure implies that every state reached while waiting belongs to the corresponding one-step stopping region. Thus, $w(S_k)\geq0$ and $w(S_k)\geq\d E[w(S_{k+1})\mid S_k]$ along any continuation path, implying that $\{\d^{k-t}w(S_k)\}_{k\geq t}$ is a nonnegative supermartingale. Iterated expectations for bounded stopping times, followed when necessary by passage to the limit under the maintained integrability conditions, show that no later stopping rule yields more than $w(s)$. No sale yields zero, which is also no larger than $w(s)$. Thus, immediate stopping is optimal on $B_t$. If $s \notin B_t$, the same argument as in the finite-horizon case shows that continuation is better than immediate stopping. Therefore, the 1-SLA rule selects an optimal action at every state and is optimal. \endpf

\pfo{Proposition 4} Suppose first that the horizon is finite. In period $T$,
$$V_T(x,r)=\max\{x-r,0\}=W_T(x-r).$$
Suppose that $V_{t+1}(x,r)=W_{t+1}(x-r)$. Conditional on the current state $(x,r)$, the next state is
$$
 (Y,R)=\bigl(\g x+(1-\g)\e_{t+1},\g r\bigr).
$$
Hence
$$C_t(x,r)=\d E[V_{t+1}(Y,R)]=\d E\left[W_{t+1}\bigl(\g(x-r)+(1-\g)\e_{t+1}\bigr)\right].$$
It follows that $V_t(x,r)=W_t(x-r)$, completing the induction. When the horizon is infinite, the same identity holds for every finite-horizon truncation and passes to the limit under the maintained convergence conditions.

When $r=0$, (23)--(24) coincide with the complete-information Bellman equation. Therefore, the threshold for the one-dimensional state $z=x-r$ is $c_t^0$. Sale is optimal if and only if
$$
 x-r\geq c_t^0,
$$
which is equivalent to (25). \endpf

\app{C: The power-transition example}

Let $T=2$, $\d=1$, $F_1(x)=x$ on $[0,1]$, and
$$F_2(y\mid x)=y^x,\qquad 0\leq y\leq1.$$
Then
$$\psi_1(x)=2x-1,
\qquad
\psi_2(x,y)=y+{1-x\over x}y\ln y.$$
The seller stops in period 2 if and only if
$$y\geq c_2(x)=\exp\{-x/(1-x)\}.$$
Hence, the period-1 continuation value is
$$\eqalign{C_1(x)
&=\int_{c_2(x)}^1\psi_2(x,y)xy^{x-1}dy\cr
&={x^2+2x-1+(1-x)\exp\{-x(1+x)/(1-x)\}\over(1+x)^2}.}$$
The seller stops in period 1 when
$$2x-1\geq C_1(x).$$
The unique crossing is approximately $.6267$. Thus, the period-1 allocation is a threshold, and after a continuation report $\hat x$ the period-2 allocation is
$$q_2(\hat x,y)=1\{y\geq c_2(\hat x)\}$$
and
$$J_1(\hat x,x)
=-\int_{c_2(\hat x)}^1y^x\ln y\,dy. \eqno(C1)$$
Since $y^x\ln y\leq0$ over the relevant region,
$$0\leq J_1(\hat x,x)
\leq-\int_0^1y^x\ln y\,dy
={1\over(1+x)^2}\leq1. \eqno(C2)$$
Moreover, $c_2(\hat x)$ is decreasing in $\hat x$, so a higher continuation report expands the domain of integration in (C1). Hence $J_1$ is increasing in the continuation report. Proposition 2 establishes implementability.

On the other hand,
$$I_2(y\mid x)=-{y\ln y\over x},$$
which can exceed one for small $x$. Thus condition (21) is not necessary. Although the current type can affect the next-period type by more than one at some realizations, the expression relevant for incentive compatibility weights this effect by the period-2 allocation rule and integrates it over the next-period type distribution. Because allocation occurs only when the next-period type exceeds $c_2(\hat x)$, this expression can remain bounded by one and increasing in the report, as (C1) and (C2) show.

\ref

\paper{Akan, M., Ata, B., and Dana, J. D.}{2015}{Revenue management by sequential screening}{{\it Journal of Economic Theory\/} 159, Part B}{728--774}

\paper{Arve, M., and Zwart, G.}{2023}{Optimal procurement and investment in new technologies under uncertainty}{{\it Journal of Economic Dynamics and Control\/} 147}{Article 104605}

\paper{Ashlagi, I., Daskalakis, C., and Haghpanah, N.}{2023}{Sequential mechanisms with ex post individual rationality}{{\it Operations Research\/} 71(1)}{245--258}

\paper{Battaglini, M.}{2005}{Long-term contracting with Markovian consumers}{{\it American Economic Review\/} 95(3)}{637--658}


\paper{Bergemann, D., and Strack, P.}{2015}{Dynamic revenue maximization: A continuous-time approach}{{\it Journal of Economic Theory\/} 159, Part B}{819--853}

\paper{Bergemann, D., and V\"alim\"aki, J.}{2019}{Dynamic mechanism design: An introduction}{{\it Journal of Economic Literature\/} 57(2)}{235--274}

\paper{Besanko, D.}{1985}{Multi-period contracts between principal and agent with adverse selection}{{\it Economics Letters\/} 17(1--2)}{33--37}

\paper{Board, S.}{2007}{Selling options}{{\it Journal of Economic Theory\/} 136(1)}{324--340}

\paper{Board, S., and Skrzypacz, A.}{2016}{Revenue management with forward-looking buyers}{{\it Journal of Political Economy\/} 124(4)}{1046--1087}

\book{Chow, Y. S., Robbins, H., and Siegmund, D.}{1971}{Great Expectations: The Theory of Optimal Stopping}{Houghton Mifflin, Boston}

\paper{Courty, P., and Li, H.}{2000}{Sequential screening}{{\it Review of Economic Studies\/} 67(4)}{697--717}

\paper{Dilm\'e, F., and Garrett, D. F.}{2026}{A dynamic theory of random price discounts}{{\it Review of Economic Studies\/} 93(3)}{1671--1709}

\paper{Doval, L., and Skreta, V.}{2024}{Optimal mechanism for the sale of a durable good}{{\it Theoretical Economics\/} 19(2)}{865--915}

\wp{Ferguson, T. S.}{2006}{Optimal Stopping and Applications}{Electronic text, UCLA}

\paper{Gallien, J.}{2006}{Dynamic mechanism design for online commerce}{{\it Operations Research\/} 54(2)}{291--310}

\ve
\paper{Garrett, D. F.}{2016}{Intertemporal price discrimination: Dynamic arrivals and changing values}{\aer 106(11)}{3275--3299}

\paper{Gershkov, A., Moldovanu, B., and Strack, P.}{2018}{Revenue-maximizing mechanisms with strategic customers and unknown, Markovian demand}{{\it Management Science\/} 64(5)}{2031--2046}

\paper{Hu, Z., and Xiao, Y.}{2025}{Optimal dynamic mechanism under customer search}{{\it Operations Research\/} 73(4)}{2045--2060}

\paper{Kakade, S. M., Lobel, I., and Nazerzadeh, H.}{2013}{Optimal dynamic mechanism design and the virtual-pivot mechanism}{{\it Operations Research\/} 61(4)}{837--854}

\paper{Krasikov, I., and Lamba, R.}{2026}{On dynamic pricing}{{\it American Economic Journal: Microeconomics\/} 18(1)}{174--207}

\paper{Kruse, T., and Strack, P.}{2015}{Optimal stopping with private information}{{\it Journal of Economic Theory\/} 159, Part B}{702--727}

\paper{Lippman, S. A., and McCall, J. J.}{1976}{The economics of job search: A survey}{{\it Economic Inquiry\/} 14(2)}{155--189}

\paper{McCall, J. J.}{1965}{The economics of information and optimal stopping rules}{{\it Journal of Business\/} 38(3)}{300--317}

\paper{Mierendorff, K.}{2016}{Optimal dynamic mechanism design with deadlines}{{\it Journal of Economic Theory\/} 161}{190--222}

\paper{Myerson, R. B.}{1981}{Optimal auction design}{{\it Mathematics of Operations Research\/} 6(1)}{58--73}

\paper{Pai, M. M., and Vohra, R. V.}{2013}{Optimal dynamic auctions and simple index rules}{{\it Mathematics of Operations Research\/} 38(4)}{682--697}

\paper{Pavan, A., Segal, I., and Toikka, J.}{2014}{Dynamic mechanism design: A Myersonian approach}{{\it Econometrica\/} 82(2)}{601--653}

\book{Peskir, G., and Shiryaev, A. N.}{2006}{Optimal Stopping and Free-Boundary Problems}{Birkh\"auser, Basel}

\bye